\documentclass[a4paper,twocolumn,english,aps,prl,showpacs]{revtex4-1}
\pdfoutput=1
\usepackage[pdftex]{graphicx,color} 
\usepackage[T1]{fontenc}
\usepackage[latin1]{inputenc}
\usepackage{amsmath}
\makeatletter

\usepackage{babel}

\begin{document}

\title{Thermopower in the Quantum Hall Regime}

\author{N. d'Ambrumenil$^1$
and R.H. Morf$^2$}

\affiliation{$^1$Physics Department, University of Warwick, Coventry CV4
7AL, UK \\ 
%$^2$Physics Department, Harvard University, Cambridge, MA 02138, USA\\
$^2$Paul Scherrer Institute, CH-5232 Villigen, Switzerland
}

\date{\today}

\begin{abstract}
We consider the effect of disorder on the themopower in quantum Hall
systems. For a sample in the Corbino geometry, where dissipative
currents are not carried by edge states, we find that thermopower
behaves at high temperatures like a system with a gap and
has a maximum which increases as the temperature is reduced. At lower
temperatures this maximum reduces as a function of
temperature as  a result of tunneling across
saddle points in the background potential. Our
model assumes that the mean saddle point height varies linearly with
the deviation in filling factor from the quantized value. We test this
hypothesis against observations for the dissipative electrical
conductance as a function of temperature and field and find good
agreement with experiment around the minimum.
\end{abstract}
\maketitle

Theoretical  analysis of heat transport and thermopower in quantum
Hall systems has concentrated on
  homogeneous systems in which the response is that of free quasiparticles or
  quasiholes and in which impurities are treated (if at all) 
as point-like scatterers
\cite{CooperHalperinRuzin97,BarlasYang12,YangHalperin09}.
In such homogeneous electron fluids, the thermopower is related to the
entropy per charge \cite{CooperHalperinRuzin97,YangHalperin09}. 
This information may be hard to
extract experimentally given that the charge and heat currents are known to be
greatly affected by edge contributions which may be
different for heat and charge transport. However, measurements in the Corbino geometry 
are expected to get round many of the complications associated with
inhomogeneous current distributions \cite{BarlasYang12}.

Localized puddles of compressible fluid within the quantum Hall state
were predicted theoretically \cite{Efros88_2} and observed directly in
scanning tunneling measurements \cite{Martinetal05}. 
While the quantized Hall response is
that of the percolating fluid, the dissipative response is thought to
be dominated by the motion between compressible puddles. Excitations
in local equilibrium in one compressible
puddle are transferred through the incompressible region to a
neighboring puddle across saddle points 
in the potential. 
The average saddle point height is $\Delta_s/2$, where $\Delta_s$ is the energy to
nucleate a particle and hole near the saddle point. This energy
controls the thermally activated dissipative conductivity at the
center of a quantum Hall plateau which varies as
$e^{-\Delta_s/2T}$ in the absence of quantum tunneling
\cite{Polyakov94}. When tunneling through saddles is
important, the dissipative conductance is larger, with an apparent
activation energy smaller than $\Delta_s/2$ \cite{dABIHMorf11}. 

\begin{figure}[t]
\includegraphics[width=3in]{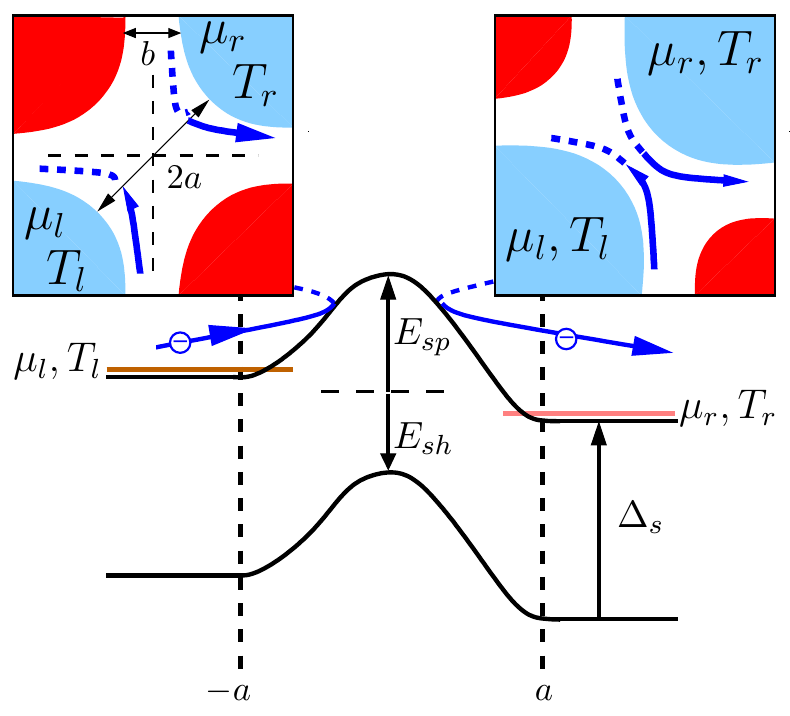}
\caption{\label{fig:bup} 
Band alignment and particle flow across a typical saddle point.
Upper left: At the minimum of the dissipative conductance
($\nu=\nu_m$, $\delta \nu=0$) the typical
saddle point heights, $E_{sp} = E_{sh}= \Delta/2$, and the saddle point
widths are equal. For $\delta \nu > 0$ (upper right) the typical
saddle points for  (blue)  qp-rich regions will grow and the saddle point width
narrow while the (red) qh-rich regions shrink. 
Transfer across the saddle is via thermally activated tunneling. The
net transfer reflects the potential and the temperature differences,
$(\mu_r-\mu_l)$ and $(T_r-T_l)$, across the saddle. In a thermopower
measurement there is no net flow. 
}

\end{figure}

Here we show that the thermopower 
is controlled by the response at saddle points (see
Fig.~\ref{fig:bup}). 
Even in the presence of
a percolating incompressible region, thermal excitations will
be predominantly located in the compressible regions where there is no
gap. The dissipative response of quantum
Hall systems is then controlled by the response of such excitations 
to temperature and voltage differences across saddles.
The link with the entropy per charge of the system, expected for
homogeneous systems, is not
simple in this context. The connection of the thermopower to entropy follows from the basic
thermodynamics of the incompressible fluid, while the dissipative response in the
inhomogeneous systems is determined by the coupling at saddle points
in the background potential of separate reservoirs of localized excitations
assumed to be in local equilibrium.

The case of transport occurring as a result of a chemical potential
potential difference (voltage drop) was treated in \cite{Polyakov94} 
without taking account of tunneling, which was then discussed in \cite{dABIHMorf11}. 
A temperature difference across a saddle point would also lead to a
flow of charge. The rate of transfer of quasiparticles with
charge, $-qe$,
across a saddle, shown moving left to right in Fig \ref{fig:bup}, is
\begin{equation}
i_{lr} =  \frac{1}{h}  \int_{0}^{\Delta_s} dE\;  {\cal T} (E-E_{sp}) e^{- (E-\mu_l) /
T_l}.
\label{eq:transfer}
\end{equation}
Here $T_l$ and $\mu_l$ are the temperature and chemical potential in the left puddle, $\cal T$ is the
transmission probability for a particle across the saddle and $E_{sp}$ is
the height of the saddle for quasiparticles. $E_{sp}$ 
can vary between $0$ and $\Delta_s$, where
$\Delta_s$ is the saddle point gap. Excitations with energy
$\Delta_s$ (or higher) would correspond to qp's having crossed the
incompressible fluid and
localizing as a `minority' carrier above the neighboring qh-rich
region (see Fig \ref{fig:bup}). 
Such processes would imply a connection between qp- and qh- channels and
would need to be taken into account at high temperatures.

When the incompressible region is wide and there is negligible tunneling across
the saddle, excitations only cross when $E>E_{sp}$ so that ${\cal
T}=\theta(E-E_{sp})$, where $\theta$ is the Heaviside function. In this limit, the integral in
(\ref{eq:transfer}) can be evaluated analytically. If
$\Delta_s \gg T_l$, the result is
$i_{lr} = \frac{T_l}{h} \exp{(-E_{sp}/T_l + \mu_l/T_l)}$. 
Taking $\mu_l=-\mu_r=(-qe)\delta V/2$ and $T=(T_l+T_r)/2$, the net (number)
current of quasiparticles crossing the saddle is
\begin{equation}
i_{s} = \frac{1}{h}\left( -(qe) \delta V + \left(1+ \frac{E_{sp}}{T}
  \right) \delta T \right) e^{-E_{sp}/T}.
\label{eq:PSnetcurrents}
\end{equation}

We will assume that the
saddle point heights are randomly and symmetrically distributed
about the mean, and that the net electrical and thermal
conductivity of the network from quasiparticle transport is given by taking the average saddle point
height, $E^{av}_{sp}$. As shown in \cite{Dykhne71}, this is always
the case for any network in which the $\log$ of the conductivity
(thermal or electrical) is evenly distributed about the mean. We will
come back to this point at the end.
Adding the contribution of the
corresponding network for quasiholes, which have charge
$+qe$, the electric current is related to $\delta V$ and $\delta T$ via
\begin{equation}
I =  L^{(11)}  \delta V -  L^{(12)} \frac{\delta T}{T}.
\label{eq:currentTotal}
\end{equation}
In the Corbino geometry the quantities, 
$L^{(11)}$ and $L^{(12)}$, are scalars, as the edge currents and bulk Hall current make
no net contribution to radial transport, and are given by
 \begin{eqnarray}
L^{(11)} &  = & \frac{(qe)^2}{h} (e^{-E^{av}_{sp}/T} + e^{-E^{av}_{sh}/T} )
\nonumber \\
\frac{L^{(12)}}{T} & = &  \frac{(qe)}{h} \left[
  \left(1+\frac{E^{av}_{sp}}{T} \right) e^{-E^{av}_{sp}/T}  \right.\nonumber \\
& & \left. -  \left( 1+  \frac{E^{av}_{sh}}{T} \right) e^{-E^{av}_{sh}/T} \right].
\label{eq:TransportCoeff}
\end{eqnarray}
The thermopower, $Q$, is minus the ratio of the voltage drop to the
temperature difference when there is no net transfer of charge from one
edge to the other. In units of $k_B/qe$ this gives 
$Q = - \frac1T \frac{L^{(12)}}{L^{(11)}}$, and 
\begin{equation}
Q = - \frac{ 
 \left(1+ \frac{E^{av}_{sp}}{T} \right) e^{-E^{av}_{sp}/T}
-\left( 1+
    \frac{E^{av}_{sh}}{T} \right) e^{-E^{av}_{sh}/T} }
{e^{-E^{av}_{sh}/T}+e^{-E^{av}_{sp}/T}} .
\label{eq:thermopower}
\end{equation}
At the center of a quantum Hall plateau, we expect $E^{av}_{sp}=E^{av}_{sh}=\Delta_s/2$, in
which case $L^{(12)}$ and the thermopower vanish.  In this
instance there are equal quasiparticle and quasihole flows from the
hotter edge of a sample to the cooler edge and consequently no net transfer of charge.

If the filling fraction is increased (or decreased) from its value at the plateau center, the
average height of the saddle points for quasiparticles will decrease (increase)
while that for quasiholes will increase. In either case,
$L^{(12)}$ will be
non-zero. We use
the experimentally measured width of the plateau as a measure of the
range in filling fraction/magnetic field over which the saddle point
structure remains. We denote the filling fractions, at which
the Hall conductance moves off its quantized value by $\nu_m \pm
\delta \nu_m$ and assume that, at these values, the corresponding average saddle point
gap, $E^{av}_{sh}$ (for $\nu=\nu_m - \delta \nu_m$) or $E^{av}_{sp}$ (for
$\nu=\nu_m + \delta \nu_m$), vanishes. Assuming that the variation is
linear in the deviation  $g=\delta \nu/\delta \nu_m$ (we discuss this
assumption at the end), gives
\begin{equation}
E^{av}_{sh} = \left(1-g \right) \frac{
  \Delta_s}{2} \hspace{0.1in} \mbox{and} \hspace{0.1in} E^{av}_{sp} = \left(1+g\right) \frac{
  \Delta_s}{2} .
\label{eq:Delofnu}
\end{equation}

\begin{figure}[t]
\includegraphics[width=3in]{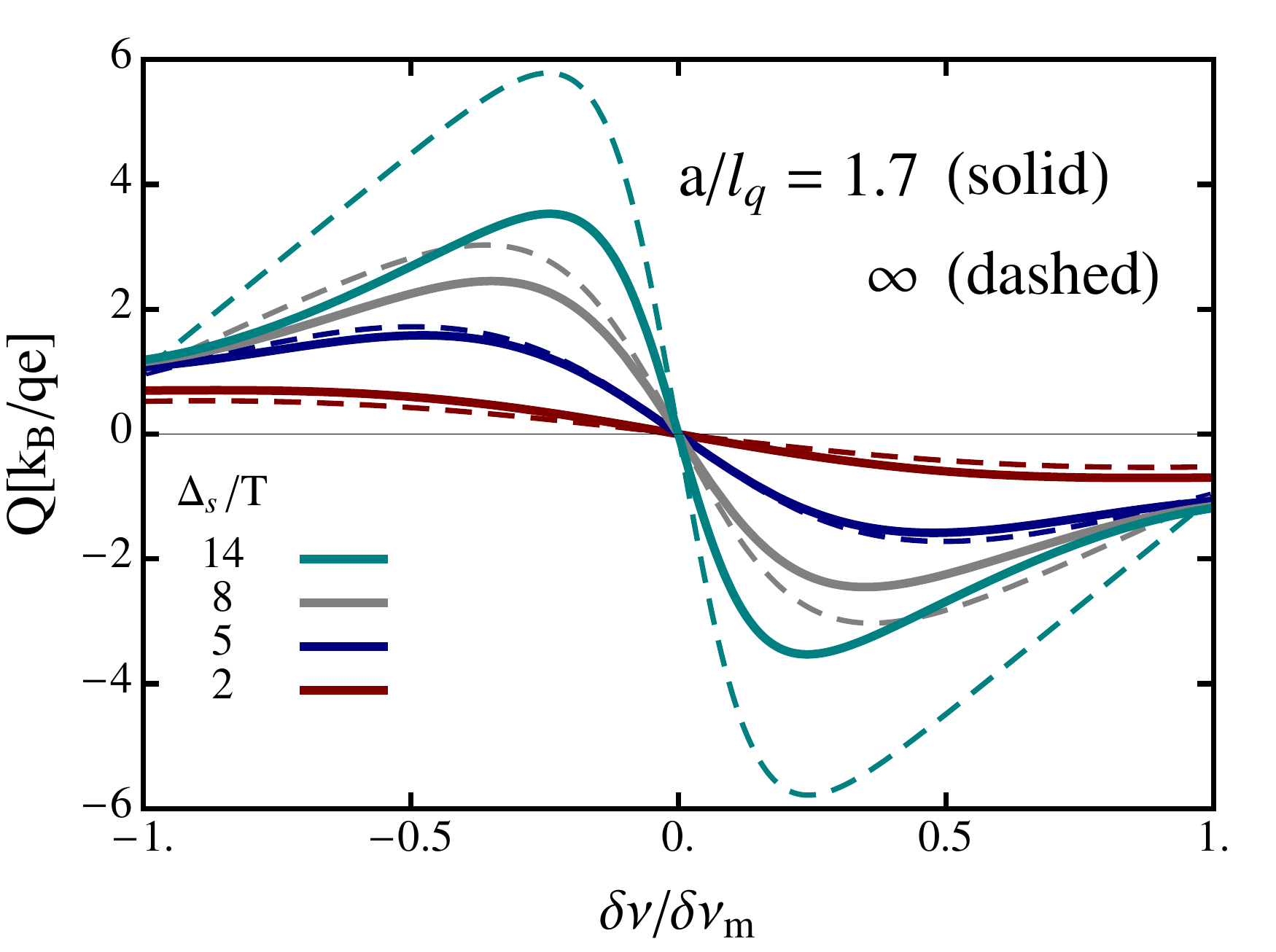}
\caption{\label{fig:TPower} 
Thermopower for the Corbino geometry in the limit $a/l_q \rightarrow
\infty$ (dashed line) and for $a/l_q=1.7$ (solid) as a function of
$\delta \nu/\delta \nu_m$ for different values of $\Delta_s/T$.
When $a/l_q\rightarrow \infty$, there is a discontinuity in $Q$ at
$g=0$. Tunneling removes this and reduces the size of the maximum/minimum. The
position of the maximum in $|Q|$ and its value are shown in Fig.~\ref{fig:Smax}.
}
\end{figure}

Combining  (\ref{eq:thermopower}) and (\ref{eq:Delofnu}) we obtain
a prediction for how the thermopower depends on temperature and
magnetic field in the limit that ${\cal T} =
\theta(E-E^{av}_{sp})$. Results
are shown by the dashed lines in Fig.~\ref{fig:TPower}. As the temperature decreases
the thermopower tends to 
$Q \approx -1 - (1-g) \Delta_s/2T$  for $g>0$ and  $Q \approx 1+(1+g)
\Delta_s/2T$ for  $g<0$. Close to $\delta \nu = 0$,  $Q$  changes by
$\Delta_s/T$   over a range in filling fractions $\delta
\nu \sim \delta \nu_m T / \Delta_s$ as $T\rightarrow 0$. A similar
result has been obtained for a homogeneous integer quantum Hall system
treated in the self-consistent Born approximation. The role of
$\Delta_s$ is played by the cyclotron
energy reduced by a Landau level broadening  parameter \cite{BarlasYang12}.

In most  samples, particularly at filling fractions with weak quantum
Hall states, the saddle point regions are not significantly wider than $l_q$ and
tunneling is important. For the special case of non-interacting
particles, with potential energy near the saddle  $W=E_s -U_x x^2 + U_y
y^2$, the transmission probability is given by  \cite{Fertig_Halperin87}
\begin{equation} 
{\cal T}(E-E_s) = 1/ (1 + e^{- \pi (E-E_s)/(\ell_q^2 \sqrt{U_x U_y} )}
).
\label{eq:TofE}
\end{equation} 
This is provided that 
$U_{x,y}/m\omega_c^2 \approx ( \ell_q/a_{x,y})^2/2 < 1$.
The exponent in the denominator was found to be given correctly by the WKB
approximation.  We assume that (\ref{eq:TofE}) holds in the
interacting case,  with $\Delta_s$
replacing the cylotron energy $\omega_c$, and assume that, for a given sample, the parameter $\sqrt{U_x
  U_y}=\Delta_s/a^2$ is the same for all saddles.

\begin{figure}[t]
\includegraphics[width=3in]{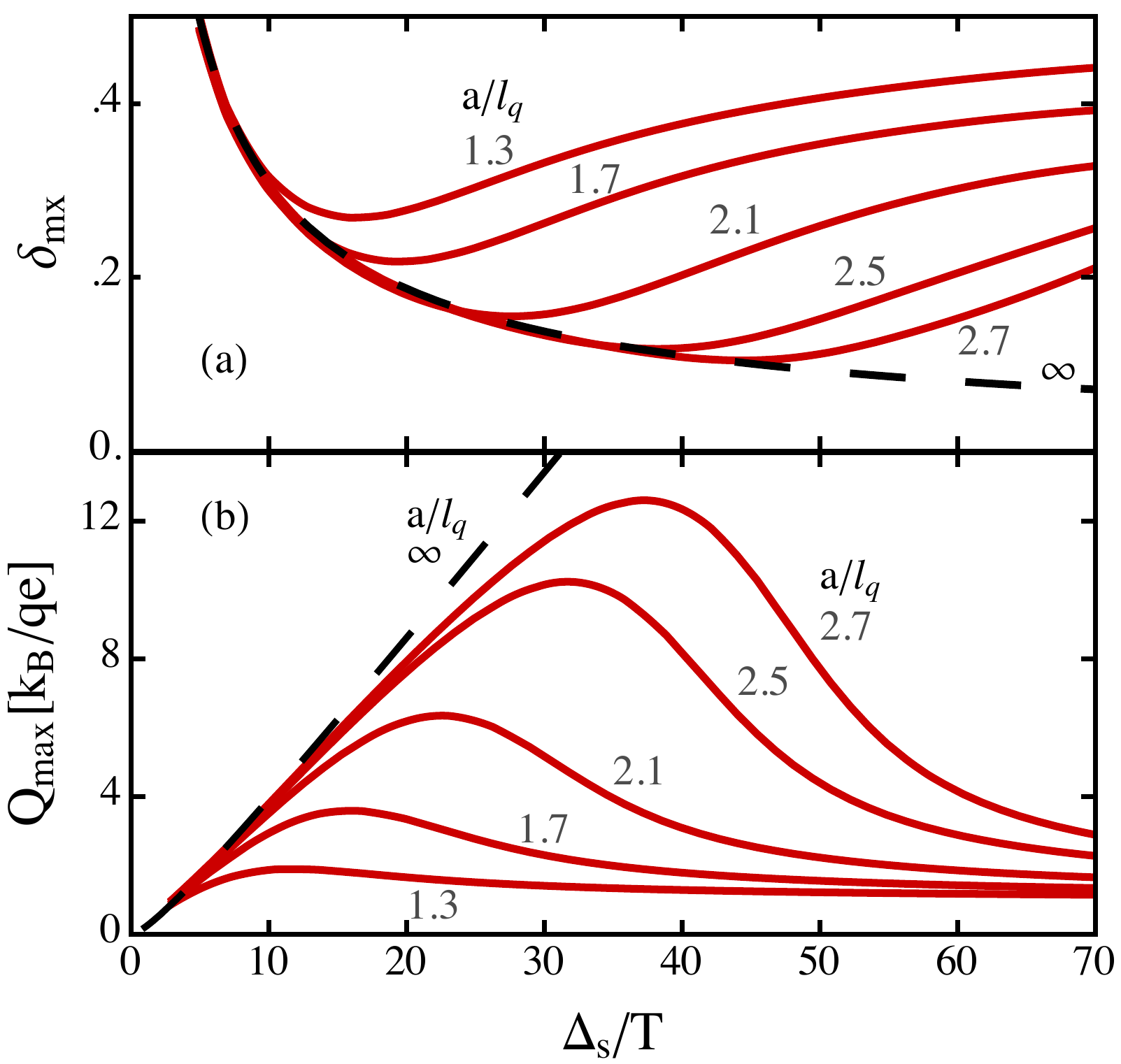}
\caption{\label{fig:Smax} 
(a) The value of
$|\delta \nu/\delta \nu_m|$ at which the maximum in $|Q|$ is observed, 
$\delta_{\mbox{mx}}$, and
(b) the maximum of  $|Q|$ 
as a function of $\Delta_s/T$. Results are shown for different values of 
the parameter $a/l_q$ (solid lines) and for $a/l_q\rightarrow \infty$
(dashed lines). }
\end{figure}

In Fig.~\ref{fig:TPower} the solid lines show how the thermopower depends on
$\delta \nu / \delta \nu_m$ for the case $a/l_q = 1.7$ for four
different values of $\Delta_s/T$. 
Decreasing the temperature leads, initially, to an increased maximum
absolute value
for $Q$ either side of $\delta \nu = 0$. The transport across a saddle
point is still thermally excited. However, with tunneling
there is significant transfer across the saddle point of qp/qh's with energies below
the saddle point and the results mimic those for a system
without tunneling but with a 
smaller gap. As the temperature is decreased, the proportion of the
current from qp/qh states with energies below the saddle point
increases. Below some temperature, which for $a/l_q=1.7$ is at $T \sim \Delta_s/15$,
 the thermopower starts to reduce. This is shown in Fig.~\ref{fig:Smax}(b),
where the maximum  of the absolute value of the
thermopower, $Q_{max}$,
as a function of $\Delta_s/T$ is plotted
for different values of $a/l_q$. In the Fig.~\ref{fig:Smax}(a), we show the value of
$\delta_{mx}$, which is the value of $\delta \nu/\delta \nu_m$ at
which the thermopower has its maximum absolute value. 

Fig.~\ref{fig:Smax}  shows a change in behavior as a function of
temperature. At high temperatures the value and
position of $Q_{max}$ is that expected for a thermally
activated system with $Q_{max}$ increasing  with decreasing $T$. 
Below a certain temperature, $Q_{max}$ 
moves away from the plateau  center, as shown in Fig.~\ref{fig:Smax}(a), 
and falls with decreasing temperatures. At very low temperatures ($\Delta_s/T\gtrsim
50$ for $a/l_q = 1.7$),  $Q_{max}$ tends towards the value
set by the entropy per charge expected for a homogeneous system.

Our model assumes that stable fractionally charged `edge excitations',
localized on the puddles, are the only objects contributing to
transport. 
However, at some filling fractions,
neutral excitations are predicted to exist which would also
be localized on each puddle. 
A microscopic study of excitations at $\nu=5/2$ suggested that the
velocity for neutral excitations, $v_n$, is an order of magnitude less
than that for charge excitations, $v_c$, for the case of a slowly varying background potential for
the underlying electrons \cite{WanHuRez08} . The barriers at saddle points in the
potential would then be substantially less for neutral than for charged
excitations and their effect on transport coefficients would become
  apparent only at much lower temperatures. If the neutral excitations are not strongly coupled to charge
excitations, there should be no contribution to conductivity
(obviously) or thermopower as the neutral excitations do not carry charge.  They would
provide an alternative channel for the transport of heat. 
Measuring the electronic contribution to all transport coefficients,
and comparing thermopower, thermal conductivity and
electrical conductivity might allow the different contributions to be
separately identified.
%If the coupling between neutral and thermal excitations were strong,
%then there might be drag effects. However, they would be small in the
%saddle point picture if the height of the saddle point
%is much higher for charge excitations than neutral excitations and the
%the dominant energy for electrical conductivity and thermopower would
%remain the saddle point gap for charge excitations.

We return now to the applicability of the model. It is valid for the temperature
regime, in which phase coherence is not established between puddles and
the excitations relax to local equilibrium in each
puddle before reaching the next saddle point. We assume
that the response of the network can then be computed from that of the
average saddle point which, if the log of the saddle point responses
is distributed evenly about a mean, is guaranteed by
Dykhne's theorem \cite{Dykhne71}. 
In the original treatment of conductivity, which does not take
account of  tunneling \cite{Polyakov94}, the conductance at a saddle
point is proportional to $e^{-E_{sp}/kT}$. As the heights of the
saddle points, $E_{sp}$,  reflect the variation of the background
impurity potential and should be evenly distributed, the
Dykhne theorem should apply.

Quantum tunneling modifies the response of the saddle so that its
log is no longer exactly symmetrically distributed about the mean for the case of
an even distribution of $E_{sp}$.
This is because, when $E_{sp,sh}
\sim 0$ %($E_{sh,sp} \sim \Delta_s$)
and the saddle point is close to disappearing, two adjacent qp/qh 
puddles are effectively shorted together by tunneling. 
In such cases the response is essentially
independent of the height of the saddle point and temperature
and it is better to think of an altered connectivity for the
two (qp and qh) networks with the formation of a larger puddle and the removal of
the saddle point. Empirically we find that the proportion of saddles,
which act as shorts and open circuits, is at most 20\% in
the case of strong tunneling ($l_q/a \lesssim 1$) \cite{dAunpub12}. Assuming that these shorts
are distributed randomly throughout the network, their contribution to
the resistance per square is negligible (the same number of `vertical' and `horizontal'
links will be affected). For the case of thermal transport, even in the absence of tunneling,
the distribution of log-conductances is not quite symmetric, 
%but
%the effects of taking account of tunneling are the same as for the
%electrical conductivity. As can be
%seen from (\ref{eq:PSnetcurrents}), the conductance varies as
%$(1+x)e^{-x}$ with $x=E_{sp}/T$, so an even distribution of $E_{sp}$
%leads to a log-conductance distribution which is biased towards small
%$x$. 
but the effect is significant only in the
cases where the saddle points are disappearing and should be accounted for via
a change in connectivity.

\begin{figure}[t]
\includegraphics[width=3.3in]{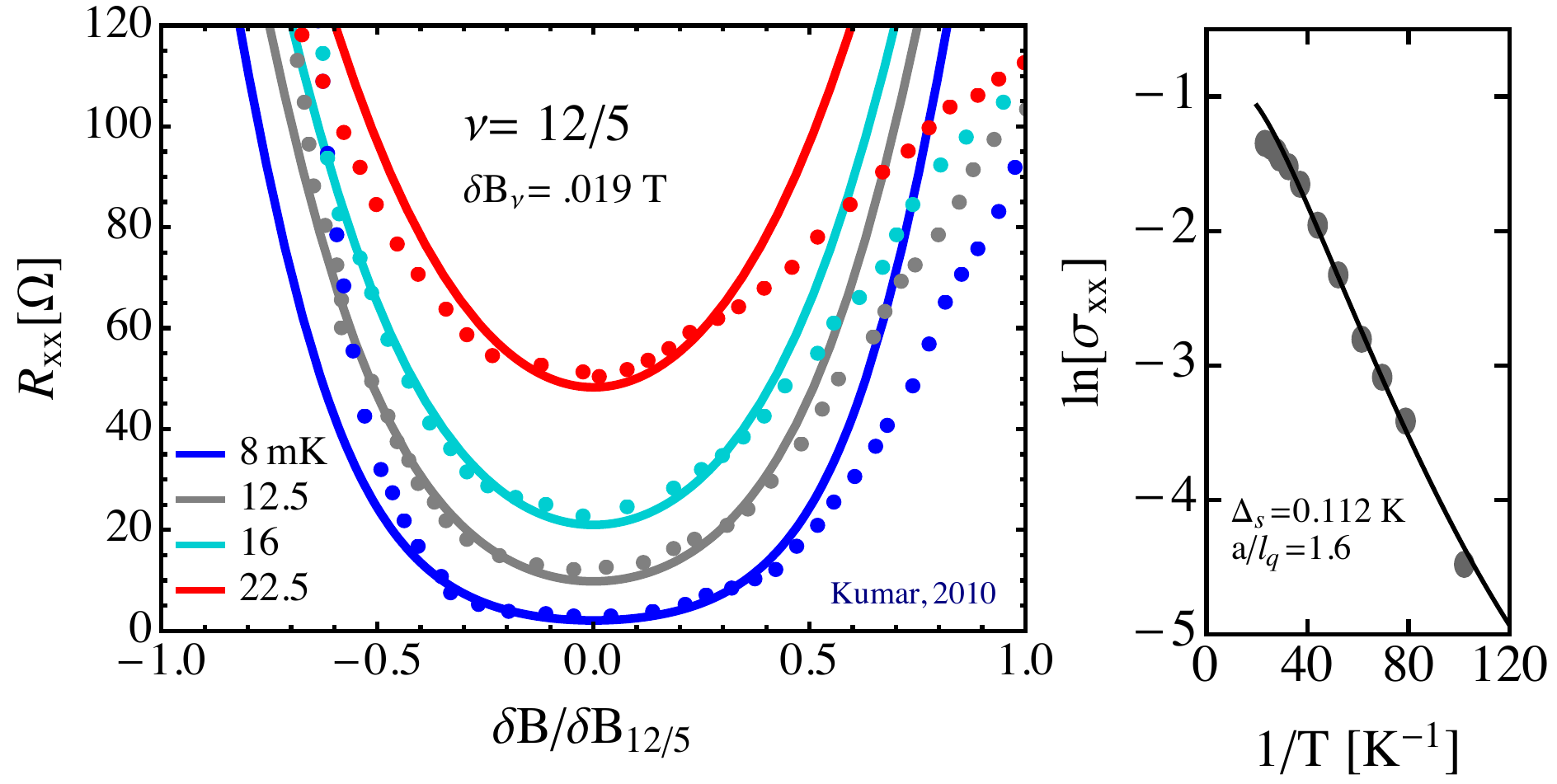}
\caption{\label{fig:KumarConduct} 
Right: Measured data points at $\nu=12/5$ from \cite{KumarCsathy10}
and the theoretical fit for $\log \sigma_{xx}$. The experimental
values for the resistance $R_{xx}$ are transformed into a conductance
measured in units of $(qe)^2/h$ with $q=1/5$.
The theoretical parameters are the saddle point gap, $\Delta_s$, the width
parameter, $a$, and the aspect ratio of the equivalent resistor
network (see text). These parameters, together with an estimate of the width of
the plateau as a function of magnetic field are used to predict the
magnetic field dependence of the longitudinal resistance at different
temperatures. The results are shown in the left panel. The data points
are digitized from the curves in \cite{KumarCsathy10}.
}

\end{figure}

We have also assumed that the average saddle point height depends linearly
on the deviation of the magnetic field from its plateau center value. We
can verify this by studying the dissipative
conductance. We convert $R_{xx}$ data into a conductance
measured in units of $2(qe)^2/h$ \cite{Polyakov94,dABIHMorf11}
 and from its temperature dependence
we estimate the saddle point gap,
$\Delta_s$,  and the width parameter, $a$. The absolute value of the
conductance then gives
an estimate of the aspect ratio of the
resistance network or sample. The aspect ratio provides a check
on the validity of the model. It should be the same for all filling fractions for a
given sample and it should not be significantly different
from one. Using the width of the plateau estimated
from the Hall resistance data, $R_{xy}$, we predict the field
dependence of the longitudinal resistance, $R_{xx}$, and compare
with experiment.  For the quantum Hall state reported in
\cite{KumarCsathy10} 
at $\nu=12/5$, we show values
of $\Delta_s$ and $a$ and the measured and predicted dissipative
conductance as a function of temperature 
in Fig \ref{fig:KumarConduct}. We assumed an aspect ratio (length to
width ratio) of 1.4 for this sample.  With these parameters and taking
the width of the plateau,
$\delta B_{12/5}$, from the traces of $R_{xy}$
we predict the
magnetic field dependence of $R_{xx}$. This is shown in the left hand
panel and compared with
experimental data points
digitized from plots in \cite{KumarCsathy10}.  The
results show that for $|\delta B/\delta B_{12/5}| \lesssim .4$ the
agreement is good. The asymmetry between positive and
negative $\delta B$ is not included in our model and presumably relates to
the nature of the excitations in the puddles which are affected by nearby
competing quantum Hall states. The prefactor
of $\sigma_{xx}$ scales with the square of the excitation charge.
At magnetic fields close to $\nu=12/5$,  excitations are likely to carry larger (smaller) charge at lower
(higher) magnetic
fields, which is consistent with what is observed.

We have shown that the
thermopower of a quantum Hall system, which is accessible in the Corbino
geometry, should be a non-monotonic function of both temperature and magnetic field close
to the plateau center. It
switches from thermally activated transport with an increasing maximum
approaching the plateau center as the temperature decreases to a
system dominated by tunneling. As the temperature decreases in this regime,
 the thermopower reduces
slowly to the value expected of a homogeneous system set by the entropy per charge.
This effect is a consequence of the
slowly varying background potential induced by ionized donors set back
from the 2DEG, and is intrinsic to any
heterostructure. We have validated our model by studying  
the electrical response as a function of magnetic field. We find
results for the dissipative conductance which agree
quantitatively with experiment for
$|\delta B/\delta B_{12/5}|\lesssim 0.4$.

\begin{acknowledgments}
We would like to thank B.I. Halperin for helpful discussions.
\end{acknowledgments}
\bibliographystyle{apsrev4-1}
\bibliography{Heat_disorder}
\end{document}